\documentstyle[epsfig]{aipproc}
\begin{document}

\def\x{{\bf x}}
\def\k{{\bf k}}
\def\H{{\cal H}}
\def\ba{\begin{eqnarray}}
\def\ea{\end{eqnarray}}
\def\be{\begin{equation}}
\def\ee{\end{equation}}
\def\ob{{\Omega _{b,0}}}
\def\oc{{\Omega _{c,0}}}
\def\tr{{\rm tr}}
\title{BEFORE INFLATION}
\author{Neil Turok}
\address{DAMTP, Centre for Mathematical Sciences, University of Cambridge\\
Wilberforce Road, Cambridge CB3 0WA, United Kingdom}

\maketitle
\begin{abstract}

Observations of the cosmic microwave sky are
revealing the primordial
non-uniformities from which all structure in the Universe
grew. The only known physical mechanism for generating the
inhomogeneities we see involves the amplification of quantum 
fluctuations during a period of inflation.
Developing the theory further will require progress in
quantum cosmology,  connecting
inflation to a theory of the initial state of the Universe. 
I discuss recent work within the framework of the
Euclidean no boundary proposal, specifically 
classical instanton solutions and the computation of fluctuations
around them. Within this framework, and for a generic inflationary
theory,  it appears that an additional anthropic constraint
is required to explain the observed Universe.
I outline an attempt to impose such a constraint in a precise mathematical 
manner.

\end{abstract}

\section{Introduction}

The maps of the cosmic microwave sky provided by recent
experiments  \cite{1}
signal an important breakthrough  in
cosmology. The structure revealed
provides our most powerful
probe of the structure in the Universe
at early times, and of its current geometry.
The data is consistent the Universe  being
nearly spatially flat, with a scale invariant 
spectrum of Gaussian-distributed irregularities in which the
overall density fluctuates but the equation of state does not.
This simplest form was 
anticipated on phenomenological
grounds as far back as the sixties \cite{2}, but later
emerged as the
prediction
of simple models of inflation \cite{3,4,5,6}.
The 
detailed agreement between theory and observations lends 
impressive
support to the view that the Universe
is, in some respects at least, remarkably simple. 

In inflationary theories, inhomogeneities 
of the required form arise a side-effect
of the exponential expansion invoked to explain the
size and flatness of the Universe. Quantum mechanical
vacuum fluctuations in the inflaton field 
are stretched to 
large scales during inflation \cite{5,6}, later to re-enter
the Hubble radius and seed large scale structure.
This magnificent 
mechanism links quantum mechanics and
gravity to the observed structure of the Universe.
It provides a direct observational probe of quantum gravity,
albeit at leading (Gaussian) order.

The underlying theory is however only provisional, 
because 
we still have no compelling model of inflation
and because a complete theory of quantum gravity is
lacking.
Nevertheless the success of the inflationary explanation of
the structure in the cosmic microwave sky forces us to 
take very seriously the idea that 
quantum mechanics governed 
the formation of the Universe we see. This moves the
topic of quantum cosmology to centre stage.
It is a field that {\it must} be developed if we are attain 
a deeper level of understanding in cosmology.

The prevalent view is that cosmology will over the next
period be 
observation-driven. 
It is true that dramatic observational developments
are certain, and there will be a lot of phenomenology to
do. However if we mean to achieve more than simply measuring
the properties of the universe, theoretical developments 
are also crucial. The great
freedom there is in current approaches to phenomenological models 
of inflation and of the early Universe, 
combined with the relatively small number of observations
(even including the cosmic microwave sky) leads me to conclude that
if we are to get anywhere, mathematical requirements of completeness
and consistency must play an increasing role.
Their power 
is already 
convincingly demonstrated by the fact that 
not one of the existing inflationary models has been made sense of
beyond leading (Gaussian) order.

In this talk I will concentrate on one particular incompleteness
of inflationary theory.
The question is why inflation started in the first place.
Vilenkin, Linde and Guth argue that the need for
a theory of initial conditions is side-stepped
because inflation is self-sustaining
to the future (i.e. `semi-eternal')\cite{eternal}. Whilst accepting that 
some theory of the 
initial conditions is needed at some level,
the claim is that 
the details of that
theory are actually irrelevant, since there
would typically be an
infinite amount of inflation between ourselves and the beginning, 
during which details of the initial conditions would have 
been erased. In this view discussions of the initial
conditions prior to inflation are fairly unimportant,
since `almost anything' will do.
In this talk I criticise the calculational methods 
which have been used to reach this conclusion, and explain a 
different point of view according to which inflation is {\it a priori} 
very unlikely and 
in any case only a brief
episode in our past.

One can easily imagine an infinite number of possible ansatzes
for the initial conditions of the Universe. At the present
stage of development of theory, many of these
might be perfectly consistent with observation. 
But among the existing proposals, I think the 
Euclidean no boundary proposal \cite{hh} is appealing because
it is based on simple and general ideas which have a rationale 
beyond cosmology. 
In a strong sense I think it is
`the most conservative thing you can do'. Of course it
may well fail precisely because it is too
conservative. Space and time may be emergent rather
than fundamental properties. Describing
the Universe as a manifold may not be appropriate to
its early moments. Nevertheless precisely because
the no boundary proposal is {\it not} just cooked up to
make inflation work, its failings and limitations may
teach us something deeper about what is in fact required.

In this lecture I focus on one particular
approach to the no boundary proposal,
using 
a set of classical instanton solutions 
of the `no boundary' form \cite{ht,th}. Our work has 
focussed on using these to 
compute the
the complete Gaussian correlators \cite{gt,het,hht,ght}, in 
a generic inflationary model (for related work see \cite{otha,othb}).
These generic instantons are unusual in several respects.
In the 
original `Einstein frame' they exhibit a curvature singularity 
which in the Lorentzian spacetime is `naked'. Nevertheless the
Euclidean action for the solutions is finite, suggesting
they should contribute to the Euclidean path integral. Second,
there is a one parameter family of solutions, with differing
action. This is 
inconsistent with their being true
stationary points of the action, but they may nevertheless be legitimate
as `constrained instantons' which are an established tool in 
other contexts \cite{affleck}.
A second feature is that the quantum fluctuations 
about these solutions are well defined to leading order, since  
the Euclidean action 
selects a unique (Dirichlet) boundary condition at the singularity.
This allows one to make unambiguous predictions for the 
microwave anisotropies in any one of these solutions, 
which Gratton, Hertog and I have recently computed \cite{gt,het,ght}. (See
also \cite{hht}).

I then describe recent work by Kirklin, Wiseman and myself \cite{ktw}
showing how the singularity apparent in the original 
`Einstein frame' may actually be removed by a change of field
variables. Thus the singularity is really 
only a coordinate singularity on superspace. 
One still has the problem that the scalar field potential
energy diverges at the singularity, and the scalar field
equation remains ill defined there. That problem is overcome if
one defines the theory as the limit of a family of theories in 
which 
$V(\phi)$ takes a particular form as $\phi$ tends to infinity,
but approximates the actual potential at arbitrarily 
large values of $\phi$. The 
limit is insensitive to the precise form 
of the regularised potential at
large $\phi$.  The regularised construction allows
one to give an improved formulation of the constraint at the
singularity, which allows for a detailed computation of the spectrum
of homogeneous modes \cite{gtnew}. The latter, we argue,
are actually crucial to a proper
interpretion of the instantons.

In the regularised theory all
quantities appearing are finite.
The classical field equations are also satisfied everywhere
if one views the instantons 
 as topologically 
$RP^4$ rather than $S^4$. 
The same construction is applied to instantons possessing two singularities
in the Einstein frame. When the singularities are
`blown up' one obtains 
a regular manifold which is
 a four dimensional analogue of the
Klein bottle. 

Finally I discuss the problem of negative modes,
complicated  by the `conformal factor' problem
of Euclidean general relativity, due to the non-positive character of
the Euclidean action. 
In the regular variables (and with the regularised potential),
the problem of negative modes about singular instantons is
well defined, and I briefly mention some of our latest
results \cite{gtnew} indicating that the most interesting singular instantons
(describing Universes with a large amount of inflation) 
do not possess any negative modes, in contrast to the non-singular
instantons of Coleman and De Luccia, and Hawking and Moss.
I comment on how this observation might help to resolve the
`empty Universe problem', if one performs a projection 
onto Universes containing the observer.

\section{Inflation and Initial Conditions}

Inflation is at heart a simple idea. 
A cosmological constant inserted into the Friedmann equation leads to
exponential expansion of the scale factor (for a Universe which is
sufficiently flat and homogeneous). The inflating region rapidly becomes 
exponentially large and smooth, and all other forms of matter
are redshifted away. This basic point seems to have been
understood a long time ago. For example an 
illustration in Peebles' book \cite{pee}
shows Professor De Sitter blowing up a large balloon.
The subtitle states
`What however blows
up the ball? What makes the universe expand or swell up? 
That is done by the Lambda. Another answer cannot be given'. 

Guth realised that scalar
fields of the type invoked in high energy particle theories 
could provide a `temporary Lambda' of just the form 
required, leading to
a period of accelerated expansion {\it before} the standard hot big bang 
\cite{3}. 
He showed how this period of cosmic inflation 
could solve the
riddles of why the observed Universe is so large, so flat and so 
uniform on large scales. 
Subsequently Linde, and Albrecht and Steinhardt \cite{4} invented 
working models involving `slow-roll' inflation.
A scalar field $\phi$ with potential energy $V(\phi)$ behaves like
a ball rolling down a hill. The equations governing the motion of the
scalar field $\phi$ and the scale factor $a$ of the Universe are:
\be
\ddot{\phi}+ 3{\dot{a}\over a} \dot{\phi} = - V_{,\phi} 
\ee
\be
\ddot{a} = {\kappa \over 3} a {\left(-\dot{\phi}^2
+V\right)} 
\ee
\be
{\left({\dot{a} \over a}\right)^2}
= {\kappa \over 3}{\left({1\over 2} \dot{\phi}^2 +V\right)}
- {k\over a^2} 
\ee
where $k=0,\pm 1$ for flat, closed or open Universes respectively.
The line element takes the form $ds^2= -dt^2 +dr^2/(1-kr^2) +r^2
d \Omega_2^2$ where $d \Omega_2^2$ is the standard line element on $S^2$. 
Here $\kappa = 8 \pi G$ with $G$ Newtons constant. From the second
equation one sees that the condition for accelerated expansion,
i.e. inflation, is $V>\dot{\phi}^2$.

If the slope $V_{,\phi}$ is 
small, the ball takes a long time to roll and the potential $V$ acts in 
the second equation just
like a cosmological constant. But as the field rolls down, 
$V$ slowly decreases and eventually
goes to zero, with the energy in the scalar field being dumped into
a bath of excitations of all the fields in the theory. This
process is traditionally called re--heating although inflation
need not have been preceded by a hot Universe epoch.
The decay of the scalar field energy into radiation 
is responsible for the generation of the 
plasma of the hot big bang.

An amazing by-product of inflation is that it neatly
provides a mechanism
for generating the primordial 
inhomogeneities which later seed 
structure in the Universe \cite{5,6}. 
The downhill roll of the scalar field is subject to 
continual quantum mechanical fluctuations, which cause it
to vary spatially.
These fluctuations may be described as due to the Gibbons-Hawking 
`temperature' of de Sitter space \cite{gh}. The 
radius of Euclideanised de Sitter space is $H^{-1}$ and
this plays the role of a periodicity scale i.e. an inverse 
temperature $\beta=T^{-1}$ in the Euclidean
path integral. The inflaton  field 
aquires fluctuations $\delta \phi \sim H$ on the Hubble radius scale $H^{-1}$
at all times, and these fluctuations become frozen in as a comoving scale
leaves the Hubble radius. 
If the classical field rolls slowly, $H$ is nearly
constant and 
one obtains a nearly scale invariant spectrum of 
fluctuations in $\delta \phi$. 
In some parts of the Universe the field fluctuates 
downhill  and in others uphill. The comoving regions
where it fluctuates down undergo re--heating earlier, and end
up with lower density than those which remain inflating for 
longer. 
The
fluctuations in the time to re--heating $\delta t$ lead to 
density perturbations:
$\delta \rho /\rho \sim \delta t H \sim (\delta \phi/
\dot{\phi}) H \sim H^2/\dot{\phi}\sim H^3/V_{,\phi} \sim (\kappa V)^{3\over 2}/V_{,\phi}$.
For example, for a quadratic potential $m^2 \phi^2$, $\delta \rho/\rho$ 
is proportional to
$m$ and fits the observations if $m \sim 10^{-5.5} M_{Pl}$
where $M_{Pl}= (8\pi G)^{-{1\over 2}}$ is the reduced Planck mass.

In the simplest models of inflation
the fluctuations are adiabatic in character, that is, all 
particle species are perturbed in fixed ratio. This prediction 
is not
really so much a consequence of inflation but rather of the assumption that
no information survives from the inflationary epoch to the hot big bang epoch
except the
overall density perturbation. In models with more than one scalar field
one generically induces perturbations of
`isocurvature' character as well. 

So much for the successes of inflation. But let us try to be a bit
more critical. A very basic question is 

$\bullet$ Why did the scalar field start out up the hill?

In almost all treatments, the answer is simply that the inflationary
theorist concerned put it there. In false vacuum inflation,
one assumes the field was stuck in the false vacuum. In slow roll
inflation, one assumes the field started out large. In
some approaches to eternal inflation 
\cite{eternal} one considers a theory which has
a potential maximum and allows inflating domain
walls. But even there 
one has to assume at least one such domain wall was initially
present. At a fundamental level, the question is unanswered.

How far up the hill did the scalar field have to be?
The number of
inflationary efoldings is given, in the slow roll approximation,
by
\be
N_e \approx \int_0^{\phi_0} d \phi 
{V(\phi) \over M_{Pl}^2 V_{,\phi}(\phi)}
\ee
where the true vacuum is at $\phi=0$, and inflation starts at 
$\phi=\phi_0$. For a monomial potential $\phi^n$, 
one has $\phi_0 \approx M_{Pl}
\sqrt{2 n N_e}$. If we require $N_e > 60$ or so to explain
the homogeneity and isotropy of our present Hubble volume, 
we see that 
 $\phi_0$ must be substantially larger 
then $M_{Pl}$ to obtain the $60$ or so efoldings needed to
explain the homogeneity and flatness of today's Universe.

Why was the initial value of the scalar field so high? 
It may be convenient for now to simply say 
`why not?', and leave the matter there. But then I think
one has to concede that 
at some level one has simply assumed the desired result.

One attempt to answer to the question of the 
required initial conditions for inflation is the
theory of `eternal inflation', advocated by Vilenkin, Linde and
Guth and others \cite{eternal}. The 
idea is that the same quantum fluctuations 
which produce the density fluctuations  have an important 
backreaction effect upon inflation itself. 
That is, the scalar field can fluctuate
uphill as well as down, and these fluctuations can compete with the classical
rolling. Comparing the change in $\phi$ due to classical rolling 
in a Hubble time, $\delta \phi \sim H^{-2} V_{,\phi}$, with that
due to quantum fluctuations $\delta \phi \sim H$, one sees that for a
monomial potential the quantum fluctuations actually
dominate at large $\phi$. 
For example, for a quadratic $m^2 \phi^2$ potential 
normalised to COBE
this occurs when $\phi > (M_{Pl}/m)^{1\over 2} M_{Pl} \sim 10^3 M_{Pl}$. 

The scenario is that at all times there are 
some regions of the Universe in which the scalar field 
takes large values. Classical rolling down from these regions
then produces large inflated and 
reheated regions. However, in 
the high
field regions the field also fluctuates uphill. This 
process can continuously regenerate the 
high field regions. 
People have tried to describe this process using
stochastic equations which couple the quantum-driven diffusion to
the classical Friedmann equation, but there 
are 
many problems with these calculations.
First, the averaging scheme used employs a particular time slicing 
and the stochastic equations derived are not coordinate invariant. 
(Note that 
the scalar field itself {\it cannot} be used as a time coordinate,
precisely because the condition for eternal 
inflation to occur is the same as the condition for the scalar
field (classical plus quantum) to cease to be monotonic in time.)
Second, the equation ignores spatial gradients and 
does not incorporate causality properly.
Third, the treatment is not quantum mechanical. The
subtleties of quantum interference are ignored and effectively
it is assumed that the scalar field is `measured' 
in each Hubble volume every Hubble time. 

The approximate calculations
are instructive. However since they in fact 
violate every known fundamental principle of physics,
one should clearly interpret them with caution. It is interesting
however that 
even after the fairly gross approximations  made, 
a `predictability crisis' emerges which is still unresolved.
Simple potentials such as $\phi^2$ do not generally lead to
a stationary state, since the field is driven to the Planck density
where the theory breaks down. A more basic problem is that
no way is known to predict
the relative probabilities for a discrete set of 
scalar field vacua. 
This suggests some profound principle is missing, and I shall suggest
below what that principle is.

The spacetime found in the simulations has an infinite
number of
infinite inflating open `bubble Universes' (much as in the scenario
of open inflation \cite{bgt}). The source of the `predictability crisis'
is the problem of trying to 
decide how probable it is for us to be in a region of one type
or of the other, when there are infinite volumes of each.
In other words, 

$\bullet$ Where are we in the infinite `multiverse'?

This is
an infrared (large-scale) catastrophe which is unlikely to
go away, and which I think makes it very unlikely that
a well defined probability measure
will emerge from a `gods-eye' view in which
one attempts to infer probabilities from an inflating 
spacetime of infinite extent.

It seems to me that one is asking the wrong question in 
these calculations. The solution may be instead to
concentrate on {\it observable} predictions. 
Theory should provide 
a procedure for calculating cosmological 
correlators like:
\be
\langle H_0^m \Omega_0^n ({\delta \rho \over \rho})^p \rangle, 
\qquad m,n,p \quad \epsilon \quad Z
\ee
where we compute the full quantum correlator and 
then take the classical part to compare with observations. 
(See 
e.g. \cite{mermin} for a discussion of this interpretation
of quantum mechanics).
We should demand the calculational procedure respects
coordinate invariance,
causality, and unitarity. Otherwise we shall be 
inconsistent with 
general relativity, special relativity or quantum mechanics.
As in statistical physics, the role of quantum mechanics is 
to provide a discrete measure. Causality, I shall argue is
equally important since it provides an infrared cutoff.

Causality is built into
special and general relativity, and is equally
present in 
quantum field theory. In a fixed background the latter
is 
perfectly causal in the sense that correlators
in a given spatial patch are completely determined by the
set of correlators on a spatial region  crossing the
the past light cone of the original patch. 
This may be seen from the Heisenberg equations of motion.
If we only ask questions 
about what is actually observable on or within our past light cone, 
we avoid the problem of dealing with the infinite number of
infinite open Universes encountered in eternal inflation, just
because the bubbles grow at the speed of light, so if you can see a bubble,
you must be inside it, and you cannot see other bubbles. 
Causality indicates it should 
be possible to define all correlators of interest in a way that
never mentions the other bubbles. 
The
question becomes not `{\it Where are we?}', in some infinite spacetime
which is pre-computed for an infinite amount of time,
but rather `{\it What is the probability for the observed Universe
to be in a given state?}' To calculate that, in a sum over 
histories approach we
need to sum over the different
possible four-Universes which could 
constitute our past. In the 
no boundary proposal, crucially, this sum is over {\it compact} Universes
bounded by our past light cone. I believe this framework
resolves the infrared problem in approaches
to eternal inflation which I mentioned above.

\section{The No Boundary Proposal}

The no boundary proposal links geometry to complex analysis and
to statistical physics. The idea is 
to contemplate four-geometries in which
the real Lorentzian Universe (i.e. with signature $-+++$) is rounded off
on a compact Euclidean four-manifold (i.e. with signature $++++$)
with no boundary. This construction can in principle
remove the initial singularity in the hot big bang. It can
be made quantum mechanical by summing over
all geometries and matter histories
in a path integral formulation.
The `rounding off' is done in analogy with statistical physics,
where one analytically continues to imaginary time. Here too
one imagines computing correlators of interest in the Euclidean
region, as a function of imaginary time, and then analytically
continuing to the Lorentzian time where correlators of
observables are needed. 

There are many technical
obstacles to be overcome in the  implementation of this approach.
These are not minor problems, and each is potentially
fatal. Until they are resolved 
doubts must remain. Nevertheless I believe progress can be made.
What I find most attractive is that the mathematical analogy
at the heart of the Euclidean proposal is perhaps the
deepest fact we know about quantum field theory, where
Euclideanisation
is the principal 
route to describing non-perturbative phenomena and
underlies most rigorous results. 
There is also an analogy 
with string theory, the theory of two dimensional quantum geometry,
where at least at 
a formal level  the 
the perturbation expansion is defined as a sum over Riemann 
surfaces, embedded 
in a Riemannian target space. Scattering amplitudes are
given 
as functions of Euclidean target space momenta, then
analytically continued to
real Lorentzian values. Euclidean quantum cosmology
follows the same philosophy.

Before proceeding let me list a few of the `technical'
 difficulties to be faced:

$\bullet$ Einstein gravity is non-renormalisable. This objection relates
to the bad `ultraviolet' properties of the
theory, which are certainly important but
 are not central to the discussion here. 
Theories usuch as supergravity with improved ultra-violet properties
are not
conceptually different as far as the problems we are discussing.

$\bullet$ The Euclidean Einstein action is not positive definite, and
therefore the Euclidean path integral is ill defined. This is the 
`conformal factor' problem in Euclidean quantum gravity
and I shall return to it below. Recent work has shown that at least
for some choices of physical variables, and to quadratic order,
this problem is overcome\cite{klt,lav,gtnew}. 
I shall mention this briefly below.

$\bullet$ The sum over topologies in four dimensions is likely
to diverge, as happens in string theory. Most likely one will
need some formulation in which manifolds of differing topologies
are treated together. 

These problems are formidable. The main hope, it seems to me,
is that 
our Universe appears to be 
astonishingly simple, well described by
a classical solution of great symmetry with small fluctuations
present at a level of a part in a hundred thousand.
This suggests that we may be able to accurately describe it
using
perturbation theory about classical solutions.

\section{Generic Instantons}

We seek solutions to the Einstein-scalar field equations which describe
a background spacetime taking the required Lorentzian-Euclidean form.
The main requirement is that there exist a special three-manifold
where the normal derivatives of the scalar field and the 
spatial metric (more precisely the trace of the second fundamental
form) vanish. If $t$ is the normal coordinate, then if this 
condition is fulfilled, the classical solution will be real
in both the Euclidean and the Lorentzian regions. Consider
the simple case where $V$ is constant and 
positive so that $V_{\phi}=0$. Then 
there is a one parameter family of classical solutions labelled by the value of
the scalar field, and in which 
scalar field is constant. The metric is that
for de Sitter space, which may be described globally 
in closed coordinates:
in which 
form 
\be
ds^2= -dt^2 +H^{-2} {\rm cosh}^2 (H t) d\Omega_3^2.
\ee
with $H^2 = {\kappa\over 3} V$. This metric possesses an analytic 
continuation to a Euclidean four sphere, if we set 
$Ht=-i({\pi \over 2} -\sigma)$. The solution possesses $O(5)$ symmetry
in the Euclidean region, $O(4,1)$ in the Lorentzian region. 
This is too much symmetry to describe our Universe, which 
only possesses the symmetries of 
homogeneity and isotropy, a six parameter group. 

However, if the potential is sloping, the maximal symmetry of the solution
is lower, only $O(4)$ in the Euclidean region. The Euclidean metric
is given by
\be
ds^2= d\sigma^2 +b^2(\sigma) d \Omega_3^2 
\ee
where $d \Omega_3^2$ is the round metric on $S^3$. 
The Euclidean Einstein equations are 
\be
\phi''+ 3{b'\over b} \phi' = - V_{,\phi} 
\ee
\be
b''= {\kappa \over 3} b {\left(\phi'^2
+V\right)} 
\ee
where primes denote derivatives with respect to $\sigma$.
If $\phi$ is constant, the second equation has solution
$b= H^{-1} {\rm sin}(H \sigma)$, describing a round $S^4$.
In general, $b$ is a deformed version of the sine function.
If the potential is gently sloping, we might expect a remnant 
of the one parameter family of solutions  to survive, 
and that is indeed the case. Consider configurations with
$O(4)$ as above, and with a regular pole which we shall take
to be  $\sigma=0$, 
where $b\sim \sigma$. If the scalar field is regular there it
must obey $\phi=\phi_0 +{1\over 2} \ddot{\phi}_0 \sigma^2 +...$,
 with $\phi_0$ an arbitrary
constant. 
If the potential is gently sloping, then $\ddot{\phi}_0$ is small 
and the 
scalar field rolls very
slowly uphill, so  
$b$ is nearly sinusoidal. However, once $\sigma$ is past the maximum
of 
$b$ the $\phi$ equation
exhibits anti-damping, and $\phi$ rolls off to infinity at some
finite value $\sigma_s$. It is not hard to show that 
$\phi$  diverges logarithmically in 
$\sigma_s-\sigma$
and $b$ vanishes as $(\sigma_s-\sigma)^{1\over 3}$.
This behaviour is generic for scalar
fields with gentle potentials. It is valid for example for 
several fields, for a 
nontrivial Kahler potential (i.e. a kinetic term $K_{ab}(\phi)\partial \phi^a \phi^b$)
or for fields with a nontrivial coupling to the Ricci scalar. 

\begin{figure} 
\centerline{\epsfig{file=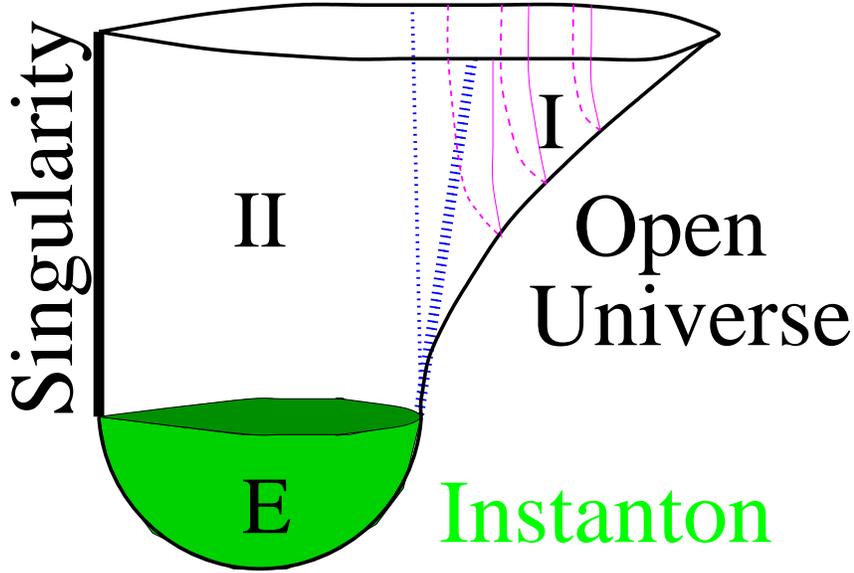,height=3.0in}}
\vskip .4in
\caption{An infinite open Universe emerges from a `pea' instanton.}
\label{fig1}
\end{figure}

As was noted in \cite{ht,th}, since all of these 
solutions have vanishing normal derivatives on `longitude' three
surfaces 
extending
from the regular to the singular pole, they may be 
may  
analytically continued to a 
Lorentzian spacetime, which takes the form
shown in Figure 1. Region I is an
infinite inflating open Universe, which has a coordinate singularity on
the lightcone through which it is connected to Region II, 
which is an approximately de Sitter region ($\phi$ is nearly constant)
bounded by a timelike singularity.
The most surprising 
result of the construction is that an infinite homogeneous
open Universe emerges from a compact Euclidean region, 
a finite object. 

So there is a one parameter family of finite action singular
solutions, in which we have a one-to-one relation
\be
\phi_0 \leftrightarrow \Omega_0
\ee
where $\Omega_0$ is the current density parameter. For the solutions
described above, $\Omega_0<1$, but if one allows instantons which
are singular at both poles, and symmetric about the maximum of $b$,
these analytically continue to closed inflating Universes, and one can 
obtain any value of $\Omega_0>1$. 

The idea of the
construction is that, in principle at least,
{\it everything} we could possibly wish to calculate
could be computed by
computing Euclidean correlators in the instanton and
analytically continuing them to the Lorentzian Universe.
This is an appealing picture.

The key characteristic 
of the singular instantons which suggests they 
might contribute to the path integral is that the Euclidean
action is finite. The scalar field diverges and 
its kinetic term yields a 
the positive logarithmic divergence to the action. But this 
is cancelled by 
a similar negative contribution from the Einstein term. 
This mechanism is 
inherently linked to the `conformal factor problem' discussed above,
and it is natural to seek to regularise the singularity via a conformal
transformation, as I discuss below. 

If one takes singular instantons seriously, they allow
one to estimate the prior probability for the inflating Universe
to begin at a given value of the scalar field $\phi_0$. 
The disappointing result of \cite{ht} was that for generic
potentials at least the most favoured 
values of $\phi_0$ are rather small and do not lead to much inflation.
The most probable Universe is then essentially empty of matter. 
An attempt was made to rescue the situation with an anthropic
argument, but even this led to an unacceptable value for
$\Omega \sim 0.01$ \cite{ht}. A different argument will be made
below, according to which a value of $\Omega$ very close
to unity is predicted.

Significant criticisms have been made of the use of singular instantons, 
and of the conclusions drawn from them \cite{vilenkin,linde}.

First, since the classical field equations break down at the
singularity, it is not clear whether they hold there. 
In fact, the action varies over the one parameter
family of solutions, so they cannot all be stationary 
points of the action. This does not  mean they do not
contribute to the path integral, nor that they cannot be
used to approximate it. But they 
must be regarded 
instead as `constrained instantons', and a suitable constraint
must be introduced. This is a relatively well developed procedure
both in quantum mechanics and in field theory \cite{affleck}.
One way of imposing a constraint will be described in 
the next section. 

Second, the presence of a singular boundary in the Lorentzian region
might lead one to 
worry that matter or radiation could leak into the spacetime
from the singularity. Equivalently, what are the correct 
boundary conditions at the singularity? Here, at least to quadratic
order in the (spatially inhomogeneous) 
fluctuations I think this worry has been convincingly
resolved. As shown in \cite{gt,het}, for scalar and
tensor perturbations, finiteness of the Euclidean action selects 
a unique (Dirichlet) boundary condition. For the spatially 
homogeneous modes,
the situation is more involved.
The required boundary condition 
is only provided by a definite regularisation of the singularity.
That proposed in \cite{ktw} and implemented in \cite{gtnew}
is described below. 

Third, Vilenkin showed in an interesting paper \cite{vilenkin}
that analogous asymptotically flat instantons exist, and he argued
these would lead to an instability of flat space towards nucleation
of singularities.
This objection was addressed
in \cite{ntflat} where it was shown that if the instantons are
constructed as constrained instantons with an appropriate 
constraint 
applied on the singular boundary, then they possess no negative 
mode and therefore do not mediate an instability. The subsequent 
work of \cite{gtnew} with a better defined constraint 
confirms this.

Vilenkin then pointed out that constrained instantons might
be placed in a `necklace' of arbitrary length, yielding 
configurations of argitrarily negative Euclidean action
which would render the Euclidean path integral meaningless.

The last two objections are addressed by the regularisation explained 
below, in which the classical field equations are satisfied
everywhere, and in which a certain constraint is applied yielding
a completely non-singular description. In this description,
`necklaces' of the form envisaged 
by Vilenkin are absent.

\section{Resolving the Singularity}

We have recently shown that it is possible to regularise the
instanton solutions in the following two steps. We first show
that the geometry of the instantons may be regularised by
`blowing up' the singularity with a $\phi$-dependent 
conformal transformation i.e. a change of coordinates on field space.
Next, we replace the scalar potential $V(\phi)$ 
by a `regularised' version, in which we deform  $V(\phi)$ at 
very large $\phi$ so that it goes to zero at $\phi=\infty$. 
Then we show that the scalar field may be rewritten in terms of
a new `twisted' field, which is forced to be zero on a special 
3-manifold. After all this we show that the `twisted' 
scalar field, and the Riemannian metric,  satisfy the field equations
everywhere. This construction renders the `singular' instantons
`regular' and makes further analysis of them well defined. 

First, we note that if we change from $\sigma$ to a coordinate 
$X=\int_\sigma^{\sigma_s} d\sigma/b(\sigma)$, the metric becomes $b^2(X) (dX^2 +d\Omega_3^2)$.
Near the singularity, one finds $b^2(X) \sim X$, {\it so the conformal 
factor has a linear zero}, in these coordinates. This suggests a simple
interpretation, which we have explored, which is that the conformal 
factor $b^2(X)$ is actually a twisted field. We shall implement this
below. 

We write the Einstein frame metric as

\be
g_{\mu \nu} (x)= \Omega^2 (x) g_{\mu \nu}^{R}(x)
\ee

where $ g_{\mu \nu}^{R} $ is a Riemannian (i.e. positive definite)
metric, and we are going to allow $\Omega^2(x)$ to possess zeros
on the manifold. 

The Einstein-scalar action becomes:
\be 
\int d^4 x \sqrt{g^R} \left(-{1\over 2 \kappa } \Omega^2 R(g^R)
 -{3 \over \kappa}
\left((\nabla \Omega)^2-{\kappa\over 6} \Omega^2 (\nabla \phi)^2\right)
 +\Omega^4 V(\phi)\right) 
\ee
 We should also add a surface term 
$-{1\over \kappa} \int_B \sqrt{h^R} \Omega^2 K^R$ if we wish the action to
involve only first derivatives of the metric.

Now the point is that $\Omega(x)$ is to be regarded as a scalar field
living on the Riemannian manifold with metric  $ g_{\mu \nu}^{R} $.
The kinetic terms for the two scalars, $\Omega(x)$ and $\phi(x)$,
definte the metric on the space of scalar fields, i.e. the superspace
metric. The metric clearly has a coordinate singularity at $\Omega=0$,
which we can remove by making the change of coordinates 
\be 
\Omega_1= \Omega {\rm cosh}(\sqrt{\kappa\over 6}\phi) \qquad 
\Omega_2=\Omega {\rm sinh}(\sqrt{\kappa\over 6}\phi).
\ee
The global structure of superspace is seen to be $1+1$ Minkowski space,
and we may change to 
light cone coordinates 
\be
\Omega_{\pm}\equiv
\Omega_1\pm \Omega_2.
\ee
This change of variables has the effect that
the action density is now finite term by term for the 
`singular' instantons. To make the whole construction analytic, 
we need to re-express the scalar field in terms of the
light cone variables, which we do via
\be
\phi= \sqrt{3\kappa \over 2} {\rm ln}\left(\Omega_+/\Omega_-\right).
\ee
But now when re-expressed in terms of $\Omega_\pm$ 
the potential $V$ has a branch cut at
$\Omega_-=0$. We therefore need to modify the potential $V(\phi)$ 
so that it is analytic as $\Omega_-$ tends to zero. Since
the term entering the action is $\Omega_+^2\Omega_-^2 V$, 
the action will be analytic in the vicinity of the conformal zero
as long as $V \sim (e^{\sqrt{2 \kappa
\over 2} \phi})^n$ at large $\phi$, with $n$ an integer $\leq 1$. 

However there is still a problem. If we continue the solutions through
the zero of $\Omega_-$, then we enter a region where the
Einstein metric channges from Euclidean to anti-Euclidean. 
We can avoid this by instead identifying antipodal points on
the three-sphere upon which $\Omega_-=0$. This produces a manifold
which is compact, and is topologically $RP^4$. Since 
$\Omega_-$ vanishes linearly (and is analytic in 
$X$) it will solve the field equations across the noncontractible 
$RP^3$ if we interpret $\Omega_-$ as being `twisted' on 
$RP^4$, an interpretation which is possible since the latter
is non-simply connected. This interpretation requires that 
the integer $n$ of the previous paragraph be odd, so that the
field equations are covariant under the symmetry $\Omega_-\rightarrow
-\Omega_-$. 

As discussed in \cite{ktw}, in this construction the `singular'
instantons are regularised and solve the appropriate classical field
equations everywhere. This is surprising at first sight, since
the solutions all have differing action. But this is allowed just
because the manifold cannot be covered by a single coordinate system
(due to its non-orientability) and the action must be defined 
by effectively introducing an `internal boundary', upon which
additional data enters. This is quite analogous to the magnetic
monopole solution on a two sphere. In that case, the
constraint that is used is the magnetic flux over the entire
$S^2$. Here instead it is the volume of the `internal boundary'
upon which the orientation flip occurs, evaluated in the 
Riemannian frame. 

We have had to go to substantial lengths to regularise the
singular instantons. But since they are now well defined as constrained 
objects, one can hope to test whether they provide good
approximations to the Euclidean path integral.

\section{Fluctuations and Negative Modes}

One way to test whether instantons provide sensible
approximations to the Euclidean path integral is to examine the
fluctuations about them. In fact the two point correlator 
for scalar and tensor perturbations which have nontrivial 
dependence on the $S^3$ coordinates on the instanton and 
which therefore yield the inhomogeneous density perturbations
in the open (or closed) universe, were computed in
\cite{gt,het,ght} and translated into 
predictions for the cosmic microwave anisotropies. (For related
work in nonsingular instantons see \cite{otha,othb}).
As mentioned above, no ambiguities emerged in these
calculations, in spite of the presence of the singularity.

The fluctuation modes which are homogeneous on the $S^3$
slices are much more subtle. A naive treatment instantly
encounters the conformal factor problem. Certain gauge invariant
fluctuation variables have negative kinetic terms, which
means there are an infinite number of negative modes. 
In recent work \cite{gtnew} (see also \cite{klt,lav}), we have shed light
on this problem, by showing that with certain choices of
physical variables, the kinetic term in the 
Euclidean action is actually
positive definite for the homogenous modes too. 
For regular instantons we find a finite number of negative modes,
always at least one. For singular instantons we find the 
Euclidean action {\it does not} uniquely select a boundary
condition, and the details of the regularisation 
therefore matter. Interestingly, we do find that for
instantons with large $\phi_0$ (therefore giving a nearly
flat Universe today), the constraint employed in the $RP^4$
regularisation explained above actually removes all negative modes.
Therefore the constrained Euclidean path integral is actually
well defined (only to Gaussian order) for these instantons.

In my view, the presence of physical negative modes is a 
serious problem for the `no-boundary' interpretation of 
cosmological instantons such as Coleman-de Luccia instantons
or Hawking-Moss instantons. It means that these instantons can
only be regarded as yielding approximate descriptions, appropriate
to 
the decay of an unstable state. They cannot be straightforwardly used
as a basis for defining the initial state of the Universe. 
The situation with `regularised' 
singular constrained 
instantons looks more promising, and deserves greater investigation.

\section{A Twisted Universe?}

The connection between topology and singular instantons is intriguing.
From one point of view, we simply wanted to regularise the instantons,
and the topology allowed us to do that. But there may be a more 
fundamental reason why nontrivial topology is important. 
Consider a scalar field theory on a circle, with a $Z_2$ symmetry
$\phi \rightarrow -\phi$, and with a 
`double well' potential. One can now choose the field
to be either `twisted' or `untwisted'. In the former case
the field must aquire a $-1$ factor as one traverses the circle,
and the topology of the configuration space (the set of $\phi(x)$)
is that of the Mobius strip. The point is that the lowest
energy state in the twisted sector is not one of the two
minima of the potential, but is instead a solitonic state
where the field has a single zero. There must therefore be 
a region of large scalar potential energy present.

It is tempting to speculate that this might provide an answer to the
question of why the inflaton field started `up the hill'. 
That is, in the Euclidean path integral there are naturally 
distinct topological sectors. If the topology is $S^4$, the Euclidean 
ground state
is the true vacuum of the theory, the stable minimum describing an
empty Universe. However, if the topology is $RP^4$, there
are instead two distinct sectors, in which the conformal factor is
respectively twisted and untwisted. In the former, as we have seen,
we get a family of singular instantons all yielding 
inflation and therefore non-empty Universes. 

I find it an intriguing idea that the twisted sector of quantum gravity
might contain a Universe filled with matter arising naturally, 
in the appropriate 
Euclidean `ground state'.

\section{Volume Factors and A-projections}

Finally, I want to deal with the question of the value of $\Omega_0$
predicted by the instanton approach \cite{tprep}. It is clear that for a generic
potential, the {\it a priori} 
most probable Universe does not have much inflation.
It could be that the inflaton model is wrong, the Euclidean path 
integral 
is wrong, the instanton solutions are wrong, or that all
three are wrong! But there is also 
another possibility, that we are just asking the wrong question. 
We should not after all be computing correlators of
cosmological observables alone, as in Eqn. (5) above.
Instead we 
whould insert an operator ${\cal P}$ corresponding to {\it projecting
onto the subset of states containing 
the particular observer who makes the observation.} 
Of course it would be nice if this insertion had no effect, so that  
the answer for the most likely Universe did not depend on whether the 
observer was in it. But we should be open to the possibility
that theory will never predict this, and all it will tell
us is about the most probable Universe we should see.

This is of course a formulation of the anthropic principle,
which seems a step backward to many physicists, from the goal 
of explaining the Universe from fundamental mathematical
principles. Indeed it is, and of course it would be nice
to 
have a theory which explained exactly what we see in the 
Universe and nothing else. But that seems unlikely since quantum mechanics
at best makes statistical predictions. It would still be nice if
the correct theory predicted only Universes very like ours. Again
that may be too much to ask, and we may be forced to pursue the
more limited goal of a theory which 
allows of all types, but within which those containing ourselves are 
in agreement with all the observations we can make.
Providing the theory satisfies other criteria - simplicity,
consistency, I think we would be happy with it. 
It seems to me that at least within the framework I have discussed
i.e. inflation in the context of a generic scalar potential 
and the Euclidean no boundary approach, this reduced goal is indeed the
best we can hope for. There is still a major challenge
to be confronted, which is to 
to properly formulate exactly what the projection
${\cal P}$ is. Due to space constraints I 
can only sketch an approach here. More details will appear 
in \cite{tprep}.

If we agree to discuss only the observable Universe, then 
it is clear that the projection ${\cal P}$ should be carried
out on our past light cone. We shall regard this as the
`observable Universe'. Then we want to impose a condition,
that there exists a certain very unlikely field configuration,
upon this past light cone.
I am really only assuming here that the particular
observer concerned 
is a very rare event - the argument would apply  equally 
for a beetle or even a paper clip!.
We do not need to know the details of this configuration.
but we shall need to assume is that it resulted in a definite
way from a particular configuration of growing mode linear
density perturbations. 
Now these linear density perturbations can
be traced back comoving in a simple way, right back to 
the primordial instanton. We therefore want to ask: How many 
ways are there of obtaining a Universe containing `me' from
one of the singular instantons. The point now is that there
is a `zero mode' present in the location of the 
past light cone relative to the perturbations on the instanton. 
When we perform the path integral we should
get a volume factor from integrating this zero mode, 
and it should be the roughly the volume of the constant
Euclidean time slices of the instanton divided by the volume 
of the comoving `special region'. If we fix the size
of the latter today, to be the spatial resolution with which
we want to identify the configuration, then the comoving 
volume on the instanton is smaller by a factor $e^{3 N_e}$ 
where $N_e$ is the number of efoldings of inflation.
Thus one expects the zero mode integration to give a
`volume factor' dependence. 

The posterior probability of obtaining a Universe
containing the observer, from an instanton solution with
scalar field value $\phi_0$ at the regular pole 
is therefore given 
by
\be
{\rm Exp}\left(-{S_E(\phi_0)\over \hbar}
 +3 {N_e}\right) 
\approx 
{\rm Exp}\left(
{24 \pi^2 M_{Pl}^4 \over \hbar V(\phi_0)}
+ 3 \int_0^{\phi_0} d \phi 
{V(\phi)\over M_{Pl}^2 V_{,\phi}(\phi)} \right)
\ee
where I have restored Planck's constant $\hbar$ and 
used the slow-roll formulae for the number
of efoldings of inflation given above.
For gentle monotonic potentials, the exponent is greatest
at very small field values,
and at very large field values, where the scalar potential $V(\phi_0)$
becomes of order the Planck density. Interestingly, there
is a minimum where the two terms compete, and this is 
precisely where the quantum fluctuations $\delta \phi \sim H\sqrt{\hbar}$ 
compete with the classical rolling $ H^{-2} V_{,\phi}$.

According to this
posterior probability, there is a high
probability for us to be in a Universe which started at the 
Planck density, had the maximum amount of slow-roll classical
inflation, and which is extremely flat today. I think this 
argument holds the prospect of solving the `predictability
crisis' mentioned in Section 2: the `volume factor' I have
alluded to is covariant and slicing-independent. Surprisingly,
it is equal in order 
of magnitude for both closed and open Universes. 

The final conclusion is in some respects disappointing. After all
there is no hope that a semi-classical calculation will be
accurate near the Planck density.
And if there has been
a very large amount of inflation, then the Universe is extremely
flat and there is not much hope of detecting effects coming
from the structure of the instanton, since perturbations of
those wavelenths are way beyond our Hubble volume today \cite{ght}.
Nevertheless the overall framework holds
some prospect of completeness, and seems to 
avoid the pitfalls of the `global view' adopted in eternal
inflation. The amount of time that inflation lasted can
be estimated, and for example 
in an $m^2 \phi^2$ potential,
the rolling time from the Planck density to
the bottom is only of order $M_{Pl} m^{-2}$, or about 
$10^{12}$ Planck times. 
Hardly eternal.

I wish to thank many physicists for
discussions of these issues, and in 
particular A. Guth, A. Linde, V. Rubakov and A. Vilenkin.
It is a special pleasure to acknowledge 
my collaborators S. Gratton, S. Hawking, T. Hertog, 
K. Kirklin and T. Wiseman.


\begin{references}

\bibitem{1} 
A. D. Miller
{\it et al.}, Astrophys.J. 524 (1999) L1;
P. de Bernardis {\it et al.},  Nature, 404, 955, (2000);
Hanany, S. {\it et al.}, astro-ph/0005123 (2000).
\bibitem{2} 
Harrison, E. R., 1970, Phys. Rev., {\bf D1}, 2726;
Zeldovich, Ya. B., 1972, M.N.R.A.S., {\bf 160}, 1. 
\bibitem{3} A. Guth, Phys. Rev {\bf D23,} 347 (1981).
\bibitem{4} A.D.  Linde, Phys. Lett. {\bf 108B,} 389 (1982);
A. Albrecht and P. Steinhardt, Phys. Rev. Lett. {\bf 48,} 1220 (1982).
\bibitem{5} 
S.W. Hawking, Phys. Lett. {\bf 115B}, 295 (1982); 
A.H. Guth and S.-Y. Pi, Phys. Rev. Lett. {\bf 49}, 1110 (1982);
A.A. Starobinsky,  Phys. Lett. {\bf 117B}, 175 (1982);
J. Bardeen, P. Steinhardt, and M. Turner, Phys. Rev. {\bf D28,} 679 (1983).
\bibitem{6} 
V.F. Mukhanov, H.A. Feldman and R.H. Brandenberger,
Phys. Rep. {\bf 215}, 203 (1992).
\bibitem{eternal} A. Vilenkin, Phys. Rev. {\bf D27}, 2848 (1983); A.D. Linde, 
Phys. Lett. {\bf B175}, 395 (1986); Phys. Lett. {\bf B327}, 208 (1994);
A.D. Linde, D.A. Linde and A. Mezhlumian, Phys. Rev.
{\bf D49}, 1783 (1994); 
V. Vanchurin, A. Vilenkin and S. Winitzki, gr-qc/9905097,
Phys. Rev. {\bf D61} 083507 (2000).
\bibitem{hh} 
J.B. Hartle and S.W. Hawking, Phys. Rev. {\bf D28} 2960 (1983).
\bibitem{ht} 
S.W. Hawking and N. Turok, Phys. Lett. {\bf B425} 25 (1998). 
\bibitem{th} 
S.W. Hawking and N. Turok, Phys. Lett. {\bf B432} 271 (1998).
\bibitem{affleck} 
I. Affleck, Nucl. Phys. {\bf B191} 429 (1981).
\bibitem{gt} 
S. Gratton and N. Turok, Phys. Rev. {\bf D 60}, 123507 (1999).
\bibitem{het} 
T. Hertog  and N. Turok, Phys. Rev. {\bf D 62} 083514 (2000), astro-ph/9903075.
\bibitem{hht} 
S.W. Hawking, T. Hertog  and N. Turok, Phys. Rev. {\bf D62} 063502 (2000). 
\bibitem{ght} 
S. Gratton, T. Hertog  and N. Turok, Phys. Rev.{\bf D62} 063501 (2000).
\bibitem{linde} 
A. Linde, gr-qc/9802038, Phys. Rev. {\bf D58}  (1998) 083514.
\bibitem{vilenkin} 
A. Vilenkin,  hep-th/9803084, Phys. Rev. {\bf D57} (1998) 7069;
gr-qc/9804051, Phys. Rev. {\bf D58} (1998) 067301; gr-qc/9812027. 
\bibitem{ktw} 
K. Kirklin, N. Turok and T. Wiseman, hep-th/0005062, submitted
to PRD.
\bibitem{ntflat} 
N. Turok, Phys. Lett. {\bf B458} 202 (1999). 
\bibitem{gtnew} 
S. Gratton and N. Turok, hep-th/0008235 (2000).
\bibitem{pee} 
P.J.E. Peebles, {\it Principles of Physical Cosmology}, Princeton 1993.
\bibitem{gh} 
G.W. Gibbons and S.W. Hawking, Phys. Rev. {\bf D15} 2738 (1977).
\bibitem{bgt} 
M. Bucher, A. Goldhaber and N. Turok, Phys. Rev. {\bf D52}, 3314 (1995).
\bibitem{mermin} 
N.D. Mermin, quant-ph/9801057.  
\bibitem{otha} 
J. Garriga, X. Montes, M. Sasaki and T. Tanaka, 
astro-ph/9811257, Nucl. Phys. {\bf B551} (1999) 317;
astro-ph/9706229, Nucl. Phys. {\bf B513} (1998) 343.
\bibitem{othb} 
A. Linde, M. Sasaki and T. Tanaka, Phys. Rev. {\bf D59} (1999) 123522.
\bibitem{klt} 
A. Khvedelidze, G. Lavrelashvili, T. Tanaka, gr-qc/0001041.
\bibitem{lav} 
G. Lavrelashvili, Nucl. Phys. Proc. Suppl. {\bf 88} (2000) 75.
\bibitem{tprep} 
N. Turok, in preparation (2000).

\end{references}
\end{document}